\documentclass{aa}
\usepackage{txfonts}
\usepackage{graphicx}
\usepackage{hyperref}
\usepackage{color}

\begin{document}

   \title{Fundamental properties of nearby stars \\and the consequences on $\Delta Y / \Delta Z$}

   \author{A.~A.~R. Valcarce$^{1,2}$, M. Catelan$^{2,3,4}$, J.~R. De Medeiros$^{1}$}
   \institute{
              $^1$Universidade Federal do Rio Grande do Norte, Departamento de F\'isica, 59072-970 Natal, RN, Brazil (\email{avalcarc@dfte.ufrn.br; renan@dfte.ufrn.br}) \\
	      $^2$Pontificia Universidad Cat\'olica de Chile, Centro de Astroingenier\'ia, Av. Vicu\~na Mackena 4860, 782-0436 Macul, Santiago, Chile\\ 
	      $^3$Pontificia Universidad Cat\'olica de Chile, Departamento de Astronom\'ia y Astrof\'isica, 
              Av. Vicu\~na Mackena 4860, 782-0436 Macul, Santiago, Chile (\email{mcatelan@astro.puc.cl})\\              
	      $^4$The Milky Way Millennium Nucleus, Av. Vicu\~{n}a Mackenna 4860, 782-0436, Macul, Santiago, Chile \\	      
             }
   \date{Received XXXXX XX, 2012; accepted XXXX XX, 2013}

  \abstract
   {One of the greatest difficulties in astrophysics is the determination of the fundamental stellar parameters, one of which is the initial mass fraction of helium ($Y$). However, given that $Y$ can be measured spectroscopically in only a small percentage of stars, a linear relationship is assumed between $Y$ and the mass fraction of metals ($Z$) from a canonical perspective of the chemical evolution of the galaxies. This $Y$--$Z$ relation is generally represented as $Y=Y_p +\Delta Y/\Delta Z \times Z$, with the value of the helium-to-metal enrichment ratio ($\Delta Y / \Delta Z$) assumed as a constant. However, there is no fundamental reason for every star to  have a $Y$ value on a linear scale with $Z$. Indeed, different $\Delta Y/\Delta Z$ values may be expected in different populations which have undergone different chemical enrichment histories.
   }
   {In this paper a new method for determining the fundamental stellar parameters of nearby stars is presented that uses at the same time $M_{\rm bol}$, $T_{\rm eff}$, and $\log \varg$. One of these parameters is $Y$, which is used to determine the validity of the $Y$--$Z$ relation.
   }
   {A new set of evolutionary tracks is created using the PGPUC stellar evolution code, which includes 7 masses ($0.5\le M/M_\odot\le1.1$), 7 helium abundances ($0.230 \le Y \le 0.370$), and 12 metallicities ($1.6\times 10^{-4}\le Z\le 6.0\times 10^{-2}$) for solar-scaled chemical compositions ([$\alpha/{\rm Fe}]=0.0$). The suggested method is tested using two different spectroscopic databases of nearby main sequence stars with precise parallaxes, and spectroscopic measurements of [Fe/H], $T_{\rm eff}$ and $\varg$. }
   {The proposed method is compared to other techniques used to determine the fundamental stellar parameters, where one assumes an age of 5 Gyr for all nearby stars. This comparison demonstrates that the hypothesis regarding constant age leads to an underestimation of the $Y$ value, especially for low metallicities. In addition, the suggested method is limited to masses above 0.60 $M_\odot$ and requires high-precision measurements of spectroscopic surface gravities in order to obtain reliable results. Finally, estimating masses and ages assuming a $Y$--$Z$ relation rather than a free $Y$ value may induce average errors of approximately 0.02 $M_\odot$ and 2 Gyr, respectively.
   }
   {}
  

   \authorrunning{Valcarce, Catelan \& de~Medeiros}
   \titlerunning{Fundamental properties of nearby stars, and the consequences on $\Delta Y / \Delta Z$}
   \maketitle
%

\section{Introduction}
\label{intro}

The relationship between the mass fraction of helium $Y$ and the mass fraction of metals $Z$, hereafter referred to simply as the $Y$--$Z$ relation, is of paramount importance in studies of the formation and evolution of most of the visible components of the Universe, including planets, stars, star clusters, and galaxies. That stellar evolution depends strongly on $Y$ has been well established since the 1960s, when stellar structure and evolution calculations revealed that main sequence stars with high helium abundances are brighter, hotter, and evolve more rapidly than their low-$Y$ counterparts \citep[e.g.,][]{Demarque1967, Iben_Faulkner1968, Simoda_Iben1968, Simoda_Iben1970, Demarque_etal1971, Hartwick_vandenBerg1973, Sweigart_Gross1978}. One of the most important, though often unstated, applications of the $Y$--$Z$ relation is in relation to theoretical models, in which a free parameter (the helium abundance) is avoided and it is assumed that divergences between theory and observations are only associated to differences in age, mass, and/or other free parameters, such as the metallicity. 

In its simplest (and most widely used) form \citep[e.g.,][and references therein]{Peimbert_TorresPeimbert1974, Audouze_Tinsley1976, Wilson_Rood1994}, the $Y$--$Z$ relation reads as $Y=Y_p +\Delta Y / \Delta Z \times Z$, where $Y_p$ is the primordial helium abundance \citep[e.g., $Y_p\approx0.240$;][and references therein]{Izotov_etal2007,Steigman2007,Steigman2012}, and $\Delta Y / \Delta Z$ is the helium-to-metal enrichment ratio. This relationship can also be written as $Y=Y_\odot +\Delta Y / \Delta Z \times (Z-Z_\odot)$, where $Y_\odot$ and $Z_\odot$ are the solar mass fraction of helium and metals, respectively.

The measurement of helium abundance in stars at different metallicities is a direct method for calibrating the $Y$--$Z$ relation. However, given that strong helium lines are only present in stars with effective temperatures ($T_{\rm eff}$) higher than 8\,000 K, which are only reached by main-sequence stars with masses greater than $\approx1.5\,M_\odot$, the chance of finding stars with these temperatures at low metallicities is low \citep{Suda_etal2008}, since low-metallicity stars evolve faster \citep[e.g.,][]{Simoda_Iben1968, Aizenman_etal1969, Iben1974, Simoda_Iben1970}. Exceptions are the blue horizontal branch (BHB) stars present in some globular clusters, which are old, metal-poor stars in the central helium burning phase. However, BHB stars with $T_{\rm eff}$ higher than 11\,500 K show higher metal and lower helium abundances \citep[effects that are intensified for higher $T_{\rm eff}$,][]{Grundahl_etal1999, Behr2003}, probably as a consequence of diffusion/levitation of elements. This leaves a narrow band of $T_{\rm eff}$ in which the initial helium abundance of these stars can be measured, although this remains complex due to the weakness of the lines and the high S/N required. This has been demonstrated by \citet[][]{Villanova_etal2009} in stars of NGC 6752, employing the \ion{He}{i} line at 5876 \AA. Using the same He line, \citet{Villanova_etal2009b} determined a high overabundance of helium with respect to the Sun in the hottest main sequence stars of the open cluster NGC 6475 (${\rm [Fe/H]}=+0.03$). Alternatively, based on the chromospheric \ion{He}{i} line at 10830 \AA, \citet{Dupree_etal2011} established the existence of variations in the helium abundances of red giant branch (RGB) stars of $\omega$ Cen with similar properties ($T_{\rm eff}$, brightness, and metallicity). Almost simultaneously and with the same line at 10830 \AA, \citet{Pasquini_etal2011} obtained similar results comparing two RGB stars in the globular cluster NGC~2808. Despite these significant efforts to determine helium abundances spectroscopically, present-day high-precision observations are not sufficiently accurate to study this element in detail for a large amount of stars.

Since direct He measurements are not yet available for most stars, indirect methods have been used to calibrate the $Y$--$Z$ relation semi-empirically. 
One such calibration is performed using a standard solar model and by assuming that the Sun represents all stars with solar metallicity. With this in mind, by applying the value of $Y_p$ for $Z=0$ and the solar calibration of $Y_\odot$ and $Z_\odot$, one can obtain the $\Delta Y / \Delta Z$ value. In our case, $Y_\odot=0.262$ and $Z_\odot=0.0167$ \citep{Valcarce_etal2012}, giving the result $\Delta Y / \Delta Z=1.31$. However, $Y_\odot$ and $Z_\odot$ depend on the input physics of the theoretical model \citep{Pietrinferni_etal2004, VandenBerg_etal2006, Weiss_Schlattl2008, Bertelli_etal2008, Dotter_etal2008} and the assumed solar chemical distribution \citep{Grevesse_Noels1993, Grevesse_Sauval1998, Asplund_etal2005, Asplund_etal2009, Caffau_etal2011}.

Another semi-empirical determination of the $Y$--$Z$ relation is carried out by \citet[][hereafter C07]{Casagrande_etal2007} using the sample of K dwarf stars in \citet[][hereafter C06]{Casagrande_etal2006} with known absolute bolometric magnitudes ($M_{\rm bol}$). This method involves obtaining helium abundances using isochrones in the theoretical $T_{\rm eff}-M_{\rm bol}$ plane. Assuming that all the stars in their sample are 5~Gyr old, the authors determine $\Delta Y / \Delta Z =2.1\pm0.9$ for stars with solar and above-solar metallicities. However, the helium abundances estimated in metal-poor stars by these authors are too low to be considered real. Using a similar method with K dwarf stars \citet{Gennaro_etal2010} determine a $\Delta Y / \Delta Z=5.3\pm1.4$, concluding that the assumption of a constant age for all stars in the sample leads to underestimation of $\Delta Y / \Delta Z$. This is because stars evolve faster for high helium abundances and/or low metallicities. Other studies using similar methods to determine $\Delta Y / \Delta Z$ include \citet{Faulkner1967} with $\Delta Y / \Delta Z=3.5$, \citet{Perrin_etal1977} with $\Delta Y / \Delta Z=5$, \citet{Fernandes_etal1996} with $\Delta Y / \Delta Z>2$, \citet{Pagel_Portinari1998} with $\Delta Y / \Delta Z=3\pm2$, and \citet{Jimenez_etal2003} with $\Delta Y / \Delta Z=2.1\pm0.9$. 

An additional method to determine $\Delta Y / \Delta Z$ is that presented by \citet{Renzini1994}, who determines $2<\Delta Y / \Delta Z<3$ using the ratio of horizontal branch clump stars to red giant stars in the Milky Way's bulge, compared to the theoretical prediction for different $Y$. Using a similar approach, \citet{Salaris_etal2004} determine an initial helium abundance of $Y=0.250\pm0.006$ for 57 globular clusters in a wide range of metallicities ($-2.2\lesssim[{\rm Fe/H}]\lesssim-0.3 $), that is, $\Delta Y / \Delta Z\approx0$ for globular clusters.

Of course there are other methods for determining $\Delta Y / \Delta Z$. A comprehensive review of these different methods can be found in Sect. 9 of \citet[][]{Gennaro_etal2010}. The different available methods can be summarized as follows: i)~predictions of galactic chemical evolution models ($1.5\lesssim\Delta Y / \Delta Z\lesssim2.4$), ii)~analysis of the chemical composition of planetary nebulae ($2.0\lesssim\Delta Y / \Delta Z\lesssim6.3$), and iii)~measurement of helium recombination lines of Galactic and extragalactic \ion{H}{ii} regions ($1.1\lesssim\Delta Y / \Delta Z\lesssim5.0$). However, taking into account only the last ten years, there is a general consensus favoring a $\Delta Y / \Delta Z$ value around  $\approx 2.0$.

In this paper a method for determining $\Delta Y / \Delta Z$ and the main fundamental stellar parameters of nearby field stars is presented. Section \ref{StellarModels} details the theoretical evolutionary tracks created for this purpose, which are used to obtain $\Delta Y / \Delta Z$ (Sect. \ref{Method}). Section \ref{Application} briefly demonstrates an application of this method for two databases of nearby stars, with conclusions presented in Sect. \ref{Conclusions}.


\section{Stellar models}
\label{StellarModels}

The stellar models used in this paper are calculated using the PGPUC stellar evolution code \citep[][hereafter PGPUC SEC]{Valcarce_etal2012} that was recently updated with the following input physics: radiative opacities for high \citep{Iglesias_Rogers1996} and low temperatures \citep{Ferguson_etal2005}; conductive opacities \citep{Cassisi_etal2007}; thermonuclear reaction rates \citep{Angulo_etal1999, Kunz_etal2002, Formicola_etal2004, Imbriani_etal2005}; equation of state \citep{Irwin2007}; mass loss \citep{Schroder_Cuntz2005}; and boundary conditions \citep{Catelan2007}. 

The PGPUC SEC was also calibrated using the Sun for a stellar model with a solar mass and a solar chemical composition distribution according to \citet{Grevesse_Sauval1998} to reproduce the present solar luminosity, radius, and $(Z/X)_\odot=0.0231 \pm 0.005$ ratio \citep{Grevesse_Sauval1998}. The results indicate a mixing length parameter $\alpha_l=1.896$, an initial solar helium abundance $Y_\odot=0.262$, and a global solar metallicity $Z_\odot=0.0167$. For a more detailed description see \citet{Valcarce_etal2012}\footnote{All the stellar models calculated for this paper, or those interpolated from these models (for any mass, $Y$, $Z$, or $[\alpha/{\rm Fe}]$), can be downloaded from \url{http://www2.astro.puc.cl/pgpuc}.}. 

A set of stellar models was calculated using a solar-scaled distribution of alpha-elements ($[\alpha/{\rm Fe}]=0.0$). These stellar models consider the evolution of stars from the zero-age main sequence (ZAMS) to the tip of the red giant branch. The grid of initial parameters includes i)~7 masses, between 0.5 and 1.1 $M_\odot$ at intervals of $\Delta M=0.1\,M_\odot$; ii)~7 helium abundances, for $Y=0.230$ and from $Y=0.245$ to $Y=0.370$, at intervals of $\Delta Y=0.025$; and iii)~12 global metallicities ($Z=0.00016$, 0.00028, 0.00051, 0.00093, 0.00160, 0.00280, 0.00503, 0.0089, 0.01570, 0.01666, 0.03000, and 0.06000). For this range of $Z$, iron abundances with respect to the Sun change from $-2.05\le{\rm [Fe/H]}\le0.56$ when assuming $Y=0.245$, to $-1.96\le{\rm [Fe/H]}\le0.66$ for $Y=0.370$. This variation is important for a reliable determination of the stellar parameters (see next section). 

The present study also used the set of stellar models with an enhanced distribution of alpha-elements $[\alpha/{\rm Fe}]=+0.3$ calculated in \citet{Valcarce_etal2012}. The grid of properties includes the same initial masses and helium abundances listed for $[\alpha/{\rm Fe}]=0.0$, and a set of 9 global metallicities covering $1.6\times 10^{-4}\lesssim Z\lesssim 1.57\times 10^{-2}$. 

The zero-point for $M_{\rm bol}$ was determined using the respective solar value $M_{\rm bol,\odot}=4.77$ mag \citep{Bahcall_etal1995}\footnote{A more recent $M_{\rm bol,\odot}$ value is $4.7554\pm0.0004$ mag (see Eric Mamajek's web page \href{http://www.pas.rochester.edu/~emamajek/sun.txt}{http://www.pas.rochester.edu/$\sim$emamajek/sun.txt}), which does not affect the main results, thanks to the small difference with the value considered in this work.}.

\section{Semi-empirical determination of fundamental stellar parameters}
\label{Method}

The fundamental stellar parameters (FSPs) are all the parameters of a star defining its present-day observational properties. The main FSPs include the initial mass ($M$), the initial chemical composition (usually $Z$, $Y$, and $[\alpha/{\rm Fe}]$), and Age. However, other parameters can also affect the star's present-day observational properties, including the rotational velocity, convective overshooting, magnetic field, mixing length parameter, and mass loss rate (after the main-sequence phase), among others. With this in mind, establishing the FSPs using theoretical models constrained by observational measurements is not a straightforward task. This is because the number of FSPs is generally greater than the amount of observational properties that can be measured, resulting in a degenerate mathematical problem. Thus, the present study aims to select the most important FSPs that match the number of observational constraints available. 

\begin{figure}
   \centering
      \includegraphics[width=9.0cm]{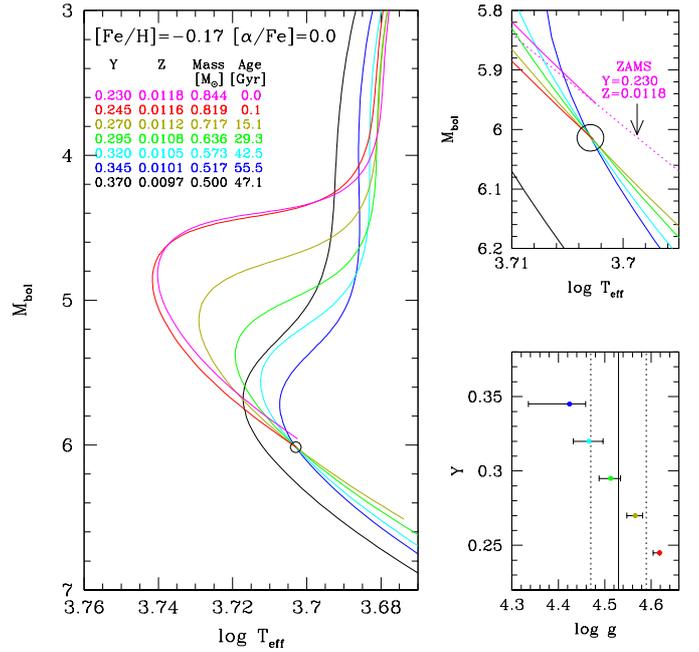}
      \caption{Method for determining the FSPs, as applied in this case to a star with $(\log T_{\rm eff}$, $M_{\rm bol})_{star}=(3.7, 6.0)$. {\bf Left panel}: First, for each star with a given [Fe/H], the masses and ages of evolutionary tracks (continuous lines) are determined for a group of helium abundances (maintaining [Fe/H] constant) at the desired point. {\bf Upper-right panel:} Zoom-in of the left panel. Next, the helium abundances with reliable matches are selected. In this case, two $Y$ values are disregarded i)~$Y=0.230$, because the star is fainter than the ZAMS for this helium abundance (dotted line), and ii)~$Y=0.370$, since a mass lower than 0.5 $M_\odot$ (the lower mass limit of the set of tracks) is required to pass through the point ($\log T_{\rm eff}$, $M_{\rm bol}$)$_{star}$. {\bf Lower-right panel}: Since $Y$ displays an almost linear dependence with $\log \varg$ at the given ($\log T_{\rm eff}$, $M_{\rm bol}$)$_{star}$ point, $Y$ can be determined using the spectroscopic $\varg$ value. The continuous and dotted vertical lines represent the spectroscopic $\log \varg$ value of the test star and its error, respectively. Dots represent the theoretical $\log \varg$ value at the point ($\log T_{\rm eff}$, $M_{\rm bol}$)$_{star}$ for each $Y$ value, and their errors are obtained assuming errors in $T_{\rm eff}$ and $M_{\rm bol}$ of $\sigma_{T_{\rm eff}}=\pm$50 K and $\sigma_{M_{\rm bol}}=\pm$ 0.05 mag, respectively.}
      \label{FigMethod}
\end{figure}

In online databases, common observational properties of nearby stars with available parallaxes are i)~$M_{\rm bol}$, or the absolute magnitude in a given filter; ii)~$T_{\rm eff}$; iii)~the spectroscopic surface gravity ($\varg_{\rm spec}$); iv)~the iron abundance with respect to the Sun ([Fe/H]); and in some cases v)~the distribution of alpha-elements with respect to the Sun ($[\alpha/{\rm Fe}]$). Using these observational properties, this paper aims to estimate $M$, Age, $Y$, and $Z$ for each star, and then obtain $\Delta Y/\Delta Z$. When [Fe/H] and $[\alpha/{\rm Fe}]$ are known, it is possible to restrict the theoretical relationship between $Z$ and $Y$ as

\begin{equation}
\label{eqFeHalpha}
{\rm [Fe/H]}_\alpha \approx {\rm [M/H]} + \log \left[\frac{ {\rm f}_{\rm Fe}(\alpha)}{\rm f_{\rm Fe}(0)} \times \frac{m_Z(0)}{m_Z(\alpha)} \right],
\end{equation}

\begin{table*}[t]
\caption{Parameters used to estimate [Fe/H].}         
\center
\begin{tabular}{lrrrrrr}
\hline
\multicolumn{1}{c}{$[\alpha/{\rm Fe}]$}&\multicolumn{1}{c}{--0.2}&\multicolumn{1}{c}{+0.0}&\multicolumn{1}{c}{+0.2}&\multicolumn{1}{c}{+0.4}&\multicolumn{1}{c}{+0.6}&\multicolumn{1}{c}{+0.8}       \\  
\hline 
\hline
\\[-1.5ex]
${\rm f}_{\rm Fe}(\alpha)$             & 0.031012     & 0.023495    & 0.016988     &0.011787      & 0.007936     & 0.005229 \\  
$m_Z(\alpha)/m_Z(0)$             & 0.989131     & 1.000000    & 1.009287     &1.016946      & 1.022616     & 1.026603 \\
$\log \left[\frac{{\rm f}_{\rm Fe}(\alpha)}
{{\rm f}_{\rm Fe}(0)} \times \frac{m_Z(0)}
{m_Z(\alpha)} \right]$           & 0.115808     & 0.000000    &-0.136819     &-0.292274      &-0.461661     &-0.641154 \\
\\[-1.5ex]
\hline
\label{TableAFe}
\end{tabular}       
\end{table*}

\noindent where ${\rm [M/H]} = \log \left(\frac{Z}{1-Y-Z}\right) - \log \left(\frac{Z}{X}\right)_\odot$, ${\rm f}_{\rm Fe}(\alpha)$ is the number fraction of iron with respect to all the elements heavier than helium; $m_Z(\alpha)$ is the average atomic mass of heavy elements weighted by the number of atoms, and ($Z/X$)$_\odot$ is the solar ratio of metals with respect to hydrogen. It is important to note that ${\rm f}_{\rm Fe}(\alpha)$ and $m_Z(\alpha)$ depend on $[\alpha /{\rm Fe}]$, and ${\rm f}_{\rm Fe}(0)$ and $m_Z(0)$ correspond to the case $[\alpha/{\rm Fe}]=0.0$. The values of the parameters that depend on $[\alpha /{\rm Fe}]$ are shown in Table \ref{TableAFe}. 

Based on the adopted reference mix, one can also obtain an expression relating the overall quantity of metals (as commonly represented by $Z$ or [M/H]) to that contained in the form of $\alpha$-capture elements and iron only (as commonly expressed in terms of $[\alpha/{\rm Fe}]$ and [Fe/H], respectively). In other words, we seek to derive the $a$ and $b$ coefficients for which the following expression is valid, for the adopted chemical composition

\begin{equation}
\label{z-zsun}
\left(\frac{Z}{Z_{\odot}}\right) = a \, \left(\frac{Z_{\alpha}}{Z_{\alpha_{\odot}}}\right) + b \, \left(\frac{Z_{\rm Fe}}{Z_{\rm Fe_{\odot}}}\right),   
\end{equation}

\noindent with $a + b = 1$, and where $Z_{\alpha}$ and $Z_{\rm Fe}$ represent the mass fractions of $\alpha$-elements and iron, respectively. Assuming that O, Ne, Mg, Si, S, Ca, and Ti are all $\alpha$-capture elements, we find $a = 0.6355$ and $b = 0.3646$. This leads naturally to an expression that is very similar to the one that was found previously by \citet{Salaris_etal1993} and \citet{Yi_etal2001}, namely 
 
\begin{eqnarray}
\label{eqFeHalpha2}
{\rm [M/H]} = {\rm [Fe/H]} + \log\big(0.6355\times f_{\alpha} + 0.3646\big),  
\end{eqnarray}

\noindent with $f_{\alpha} = 10^{[\alpha/{\rm Fe}]}$.

The first step of our method is to determine the relationship between $Z$ and $Y$ for each star with a given [Fe/H] and, indirectly, a given $[\alpha /{\rm Fe}]$. From Eq. \ref{eqFeHalpha2} it is straightforward to obtain the following relationship

\begin{equation}
Z \approx \frac{c}{1+c} \left(1-Y\right),
\end{equation}

\noindent with

\begin{equation}
c = \big(0.6355\times f_{\alpha} + 0.3646\big) \times \left(\frac{Z}{X} \right)_\odot \times 10^{[{\rm Fe/H}]}.
\end{equation}

The second step is to determine the other FSPs depending on $Y$: $Z(Y)$, $M(Y)$, and Age($Y$). To that end, a set of interpolated evolutionary tracks is used in the theoretical Hertzsprung-Russell diagram. As is shown in Fig. \ref{FigMethod}, the stellar evolutionary tracks with different $Y$ values pass through the test star at $\log T_{\rm eff}=3.7$ and $M_{\rm bol}=6.0$. However, as can be seen, there are evolutionary tracks that do not pass exactly through this point for some $Y$ values, given that i)~the star is fainter than the ZAMS for $Y=0.230$ (dotted line in the upper right panel of Fig. \ref{FigMethod}), indicating there is no mass value to solve the problem for this specific $Y$ value; or ii)~the star is cooler (hotter) than the whole evolutionary track with the minimum (maximum) mass of our database, as is the case for $Y=0.370$ in that figure. Note that an evolutionary track with a mass lower than 0.5 $M_\odot$ for $Y=0.370$ can pass through the point $(\log T_{\rm eff}, M_{bol})_{star}$, but because stars with masses lower than our limit are more difficult to detect and have large associated observational errors, these evolutionary tracks are not calculated. Note also that these stars are much more difficult to model thanks to the molecules formed in their cool atmospheres. In both cases, the FSPs for these specific $Y$ values are ignored becuase of physical inconsistency.

Finally, since each evolutionary track with a different $Y$ value has a different mass for the same $\log T_{\rm eff}$--$M_{\rm bol}$ point, the helium abundance can be determined by comparing $\varg_{\rm spec}$ with the theoretical surface gravity ($\varg$). As is known, in the $M_{\rm bol}$--$T_{\rm eff}$ diagram the differences in $\varg$ are only due to the different stellar masses, since the stellar radius is determined by the Stefan--Boltzmann law. However, when a specific filter is used instead of $M_{\rm bol}$ (for example $M_{\rm V}$), the relationship is not entirely the same since bolometric corrections do not have a linear dependency with $T_{\rm eff}$ and $\varg$. 

The bottom-right panel of Fig. \ref{FigMethod} shows the relationship between $Y$ and $\varg$ with the same $M_{\rm bol}$ and $T_{\rm eff}$ values, where each point is the value of $\log \varg$ at the test point of the track with a given $Y$ value. Note the almost linear relationship between $Y$ and $\log \varg$. Theoretical errors for $\log \varg$ are determined assuming typical errors in $T_{\rm eff}$ and $M_{\rm bol}$ of $\pm 50$ K and $\pm 0.05$ mag, respectively, based on present-day measurements of these parameters \citep[e.g.,][]{Baumann_etal2010}. Using the spectroscopic value of the test star $\log \varg=4.53\pm0.06$ dex, values represented by vertical continuous and dotted lines, and the Hermite interpolation algorithm presented by \citet{Hill1982}, the following FSPs are obtained: $Y=0.287^{+0.031}_{-0.028}$; $Z=0.0109^{+0.0005}_{-0.0004}$; $M=0.661^{+0.102}_{-0.083}\,M_\odot$; and ${\rm Age}=24.7^{+16.7}_{-16.4}$ Gyr.

Since the separation of evolutionary tracks in the $M_{\rm bol}$--$T_{\rm eff}$ diagram at the red giant branch is far smaller than the typical errors, this method must be restricted to main-sequence and sub-giant branch stars, although the latter may have a larger uncertainty in the derived age. As such, stars of the databases mentioned in the following section are restricted by $\log \varg \ge 3.8$ dex.

\subsection{Comparison of methods}

\begin{figure}
   \centering
      \includegraphics[width=9.2cm, trim=1cm 0cm 0cm 0cm]{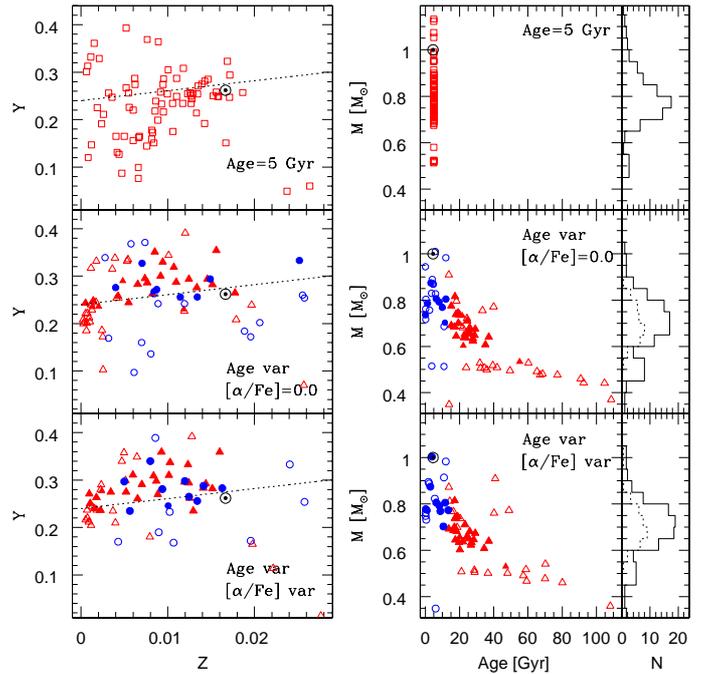}
      \caption{Helium abundance $Y$ versus metallicity $Z$ (left panels), and stellar mass versus Age (right panels) for stars listed in C06. Circles and triangles represent stars with ages younger and older than 13.5 Gyr, respectively. The Sun's properties ($Z=0.0167$, $Y=0.262$, ${\rm Age}=4.6$ Gyr, and $M=1\,M_\odot$) are represented by the symbol of the Sun. Open and filled symbols are the extrapolated and interpolated results, respectively. The dotted line represents the reference $Y$--$Z$ relation $Y=0.240+2.0\times Z$. {\bf Upper panels}: FSPs determined using the method from C07. {\bf Middle panels}: FSPs determined using the method proposed in this paper. {\bf Lower panels}: Similar to the middle panels, except theoretical evolutionary tracks have a variable $[\alpha /{\rm Fe}]$ value (see text for more details).}
      \label{FigNoExtra}
 \end{figure}

In this section the method described above is tested and compared to other similar approaches. The results are shown in Fig. \ref{FigNoExtra}, where the solar symbol and the dotted line represent the Sun's properties and the $Y$--$Z$ relation $Y=0.240+2.0\times Z$, respectively. Extrapolated results are used in this section only; the risks involved in this procedure are discussed in Sect. \ref{NoExtrap}.

Stars listed in the C06 catalog are used, consisting of nearby low-mass main sequence stars with good [Fe/H] and $T_{\rm eff}$ measurements. However, since the method suggested in this paper uses the $\varg_{\rm spec}$ value, the C06 table had to be completed with the spectroscopic values obtained by the author of the reference cited for each star. The range of iron abundances, effective temperatures, and surface gravities covered by this database is $-2.0\lesssim{\rm [Fe/H]}\lesssim+0.4$, $4400\lesssim T_{\rm eff}{\rm [K]}\lesssim6400$, and $4.1\lesssim\log \varg\lesssim5.0$, respectively, with errors of approximately $\sigma_{\rm [Fe/H]}\approx\pm 0.15$ dex, $\sigma_{T_{\rm eff}}\lesssim\pm100$ K, and $\sigma_{\log \varg}\lesssim\pm0.20$ dex, as well as some outliers. For stars without an estimated $\log \varg$ error, a $\sigma_{\log \varg}=0.05$ dex is assumed. The apparent bolometric magnitude $m_{\rm bol}$ calculated by C06 and the parallaxes for each star are also used to determine the absolute bolometric magnitude $M_{\rm bol}$.

The first method is similar to that described by C07, used to obtain the FSPs of the C06 stars. As previously mentioned, these authors assume that all stars are 5 Gyr old, and determine the helium abundance for each star using isochrones in the theoretical $M_{\rm bol}$--$T_{\rm eff}$ plane. Given that only a limited set of isochrones with different helium abundances is available in the literature, C07 use extrapolated results to obtain $Y$ values between 0.10 and 0.30. In this first approach (upper panels of Fig. \ref{FigNoExtra}), rather than obtaining the FSPs using $\varg_{\rm spec}$ (as described in the previous section), here the Age is used as the known parameter to interpolate (or extrapolate) the other FSPs for a fixed value of 5 Gyr. The results indicate that almost all stars with $Z\gtrsim0.01$ follow the adopted $Y$--$Z$ relation. However, for $Z\lesssim0.01$ the $Y$ value is significantly dispersed, with a trend towards lower $Y$ values for lower metallicities. This trend is also found by C07, but they attribute it to the current limits in stellar models.

There is an alternative explanation for the steeper trend between $Y$ and $Z$ at low metallicities. Stars that are older than 5 Gyr will be brighter than theoretical models with that age and, consequently, mistakenly appear He-depleted. This is because when $Y$ decreases the main sequence locus becomes cooler and seems brighter \citep{Demarque1967, Iben_Faulkner1968, Simoda_Iben1968, Simoda_Iben1970, Demarque_etal1971, Hartwick_vandenBerg1973, Sweigart_Gross1978}. Although this effect is not so important for stars with masses lower than 0.60 $M_\odot$\footnote{Stars with masses lower than 0.60 $M_\odot$ show almost no $M_{\rm bol}$ and $T_{\rm eff}$ variations, among other properties, over a period equivalent to the age of the Universe. Using PGPUC SEC calculations for a 0.60 $M_\odot$ star with $Z=0.01$, $Y=0.245$, and $[\alpha/{\rm Fe}]=0.0$, the variations in $M_{\rm bol}$, $T_{\rm eff}$, and $\varg$ after 13.5 Gyr are $\Delta(M_{\rm bol})\approx0.16$ mag, $\Delta(\log T_{\rm eff})\approx0.01$ dex, and $\Delta(\log \varg)\approx0.03$ dex, respectively.}, when the mass is greater than 0.60 $M_\odot$ the evolution is faster and the underestimation of helium is higher. Similarly, stars younger than 5 Gyr will appear overabundant in helium. These effects are amplified when metallicity decreases or helium increases \citep[see ][ for a recent published explanation]{Valcarce_etal2012}. As shown in the upper-right panel of Fig. \ref{FigNoExtra}, almost all stars have masses greater than $0.60\,M_\odot$, and accordingly the ages of nearby stars cannot be assumed similar to the Sun's age for comparisons of this kind.

In the middle panels of Fig. \ref{FigNoExtra} the FSPs of stars using the method proposed in the previous section are shown. In this case it is assumed that the $\alpha$-element distribution of each star was not determined, since only a small fraction of spectroscopic studies have provided estimates of $[\alpha/{\rm Fe}]$ for each star. Thus, the FSPs are obtained using evolutionary tracks with a solar-scaled distribution of elements ($[\alpha/{\rm Fe}]=0.0$). In this panel, stars older and younger than 13.5 Gyr are represented by circles and triangles, respectively. Results obtained by interpolation and extrapolation are depicted by filled and open symbols, respectively. The most notable difference from the previous approach (all stars are assumed to be 5 Gyr old) is the reduced number of stars with very low helium abundances. Moreover, there is a high percentage (68\%) of stars older than 13.5 Gyr, which is far more evident for lower masses. There are two reasons for this: i)~the current problem in predicting realistic radii for low-mass stars \citep[e.g.,][]{Torres_etal2010,Feiden_Chaboyer2012,Basu_etal2012} where errors in the radius determination of only 5\% produce differences of some tens of Gyrs, and ii)~any small error in the estimations of $M_{\rm bol}$, $T_{\rm eff}$, and/or $\varg$ can induce a large error in the age, due to the small variations expected for low-mass stars during a period equivalent to the age of the Universe. These problems induce a lower mass limit of $\sim0.60\,M_\odot$ for the reliability of estimations of $Y$ and Age with the suggested method because of the high sensitivity of these parameters to small variations of $M_{\rm bol}$, $T_{\rm eff}$, and/or $\varg$. Alternatively, masses are less sensitive to variations of these observational properties, thus making mass estimations more reliable.

The bottom panels of Fig. \ref{FigNoExtra} show the results of the most robust method, which requires the spectroscopic determination of $[\alpha/{\rm Fe}]$ for each star. This method applies the same procedure described in Sect. \ref{Method}, with a set of evolutionary tracks according to the $[\alpha/{\rm Fe}]$ value of the respective star. This set of evolutionary tracks for each $[\alpha/{\rm Fe}]$ is interpolated in accordance with the procedure described in the appendix of \citet{Valcarce_etal2012}. In order to have the same number of stars as in the middle panels, stars with $[\alpha/{\rm Fe}]<0.0$, and $[\alpha/{\rm Fe}]>+0.3$ are assumed to have $[\alpha/{\rm Fe}]$ values of 0.0, and +0.3, respectively. As demonstrated, when the $[\alpha/{\rm Fe}]$ value is taken into account, dispersion on the initial helium abundance decreases even further than in the previous case, particularly for low metallicities. Although this method is more robust than the previous one, results still indicate ages greater than the age of the Universe for stars with masses $M\lesssim 0.70\,M_\odot$. Moreover, the statistical difference between the two methods (considering or not $[\alpha/{\rm Fe}]$) is not significant, at least for this set of stars.

\subsection{Why extrapolated results should not be used}
\label{NoExtrap}

Although in some mathematical problems results can be obtained using extrapolation algorithms, these results may not always be realistic. For example, when the FSPs of stars are determined using their $\log \varg_{\rm spec}$ value (as shown in Fig. \ref{FigMethod}), there are cases when $\varg_{\rm spec}$ values are greater than $\varg$ of the minimum available $Y$ value ($Y_{\rm min}$). By extrapolating results, we can conclude that in these cases helium abundances are lower than $Y_{\rm min}$, and then obtain the other FSPs. In those cases, however, these stars must be brighter or cooler to avoid being on the wrong side of the ZAMS for the corresponding $Y$ value.

\begin{figure*}
   \centering
      \includegraphics[width=19.0cm, trim=1cm 7cm 0cm 0cm]{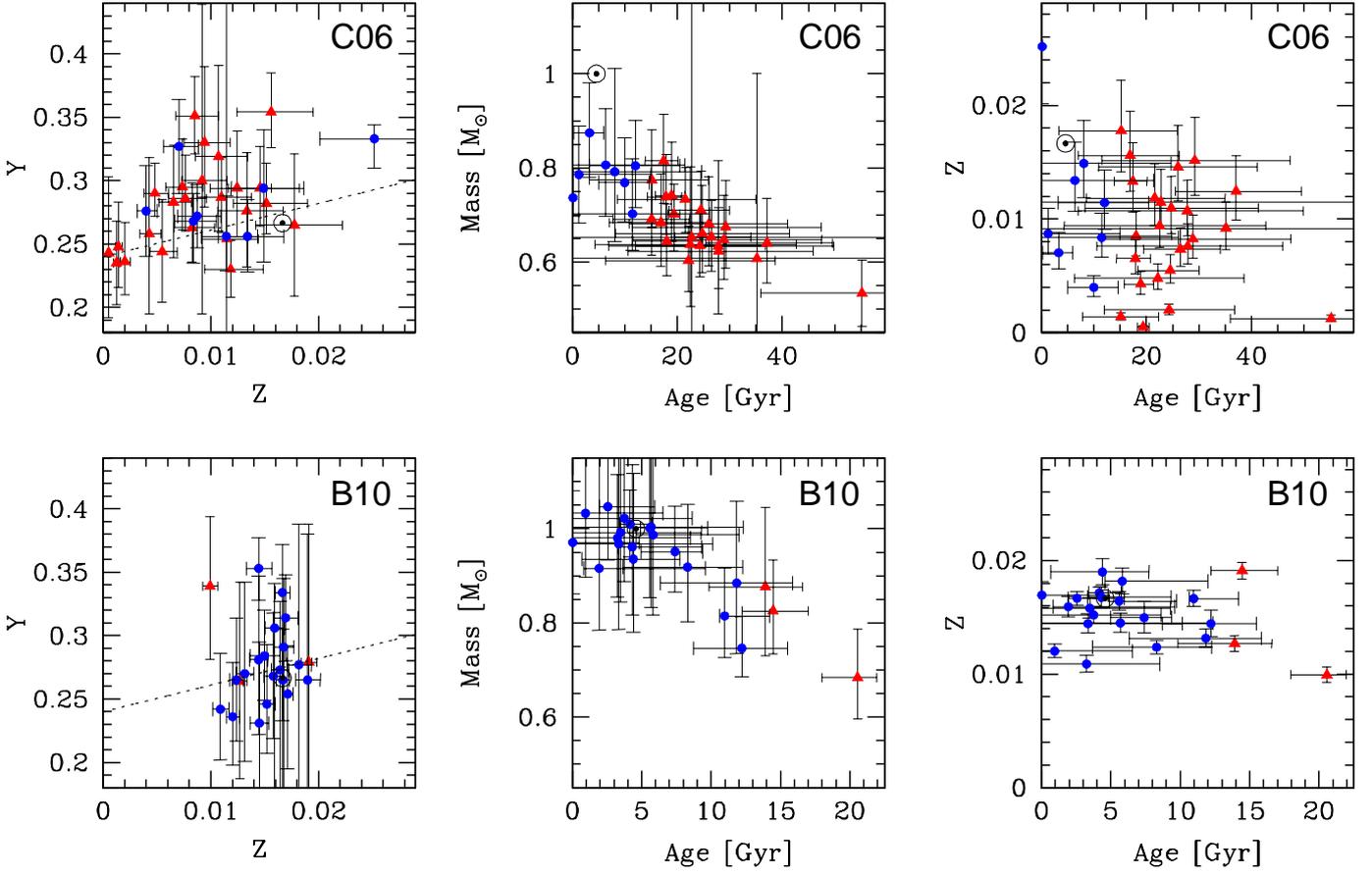}
      \caption{FSPs of stars from C06 and B10. The error bars for $Y$, $M$, and Age are determined based primarily on the error in the surface gravity ($\sigma_{\log \varg}$). The error bars in $Z$ are established by propagating the error in [Fe/H] listed by each respective author. Blue circles and red triangles indicate stars (formally) younger and older than the accepted age of the Universe ($\sim$13.5 Gyr), respectively. Dotted lines represent the $Y$--$Z$ relation $Y=0.240+2.0\times Z$.}
      \label{FigErrors}
 \end{figure*}

Since stars with extrapolated results are found on the fainter and hotter side of the ZAMS for the given chemical composition, the main findings exclude all stars with at least one FSP outside the respective range. This means that only stars with $0.23\le Y\le 0.37$, $0.5\le M/M_\odot\le 1.1$, and $1.6\times 10^{-4}\le Z\le 6.0\times 10^{-2}$ are accepted. In other words, stars with open symbols in the middle and bottom panels of Fig. \ref{FigNoExtra} are not included in the results of the next section (Figs. \ref{FigErrors} and \ref{FigB10vsV12}).

\section{Application}
\label{Application}

To apply the method proposed in this paper, one additional database is selected with different properties to those of C06. The second database was created by \citet[][hereafter B10]{Baumann_etal2010}, and consists of 117 stars with near-solar properties: $-0.4\lesssim{\rm [Fe/H]}\lesssim+0.3$, $5600\lesssim T_{\rm eff}{\rm [K]}\lesssim6100$, and $4.0\lesssim\log \varg\lesssim4.6$. These stars have excellent spectroscopic measurements, where mean errors for the spectroscopic variables are smaller than in the C06 catalog, namely $\sigma_{\rm [Fe/H]}\approx\pm 0.025$ dex, $\sigma_{T_{\rm eff}}\approx\pm40$ K, and $\sigma_{\log \varg}\approx\pm0.06$ dex. The database contains a good set of well-measured stars for comparison with the Sun. For each star, the Hipparcos parallax and $V$ magnitude are obtained from the SIMBAD database, in order to use $M_V$ as opposed to $M_{\rm bol}$. In this case, theoretical models are transformed to the observational plane using the bolometric corrections from \citet{Castelli_Kurucz2003}.

\subsection{Fundamental stellar parameters and their errors}

The FSPs for the two selected databases and their respective errors are shown in Fig. \ref{FigErrors}. Before discussing these results, it is important to note that only non-extrapolated results are being used. As such, FSPs are determined for only 33\% and 20\% of the total number of stars in the C06 and B10 databases, respectively. Using extrapolated results produces higher percent-ages, as occurs in the middle panels of Fig. \ref{FigNoExtra}, where FSPs are determined for 75\% of C06 stars. Similarly, the percentage of stars younger than 13.5 Gyr with determined results is 25\% for C06, and 87\% for B10.

Formal error bars in $Y$, $M$, and Age are established based on the error in the spectroscopic value of $\log \varg$ combined with the same analysis as in the bottom-right panel of Fig. \ref{FigMethod}. In the case of $Z$, error bars are determined by propagating directly from the error in [Fe/H]. Although we have argued against using extrapolated results, a linear extrapolation algorithm was applied to estimate error bars outside the limits of the theoretical models to make sure that errors do not seem smaller than they really are.

For the selected databases, $Y$--$Z$, $M$--Age, and $Z$--Age relationships are shown in the left, middle and right-hand panels of Fig. \ref{FigErrors}, respectively. Based on the C06 results (upper panels) one finds that the reference $Y$--$Z$ relation (dotted line) seems to fit the lower values of $Y$ for every $Z$ value, whereas the maximum $Y$ value shows a steeper increase with $Z$. In addition, no relationship is observed between $Z$ and Age. However, as previously mentioned, errors may occur in calculating $Y$ and $M$ values for stars (formally) older than the age of the Universe (red triangles) because of difficulties modeling the stellar radius. In principle, one can consider such a (formally) very old star as having underestimated $\varg$. Consequently, its helium abundance would be overestimated and mass and age underestimated. On the other hand, underestimation of its brightness (owing to parallax or the brightness itself) or overestimation of $T_{\rm eff}$ may cause overestimation of $Y$ and mass, as well as underestimation of age. As such, depending on the stellar evolutionary phase of the star and its mass the errors in the FSPs may be large or small.

For the B10 database (lower panels in Fig. \ref{FigErrors}), a substantial percentage of stars have similar metallicities, ages, and masses. However, the spread in helium abundance is greater than expected when compared to the reference $Y$--$Z$ relation. Almost all stars in this database with established FSPs are younger than 13.5 Gyr, except for three stars (or one if the error bars are considered). This illustrates the good agreement between theory and recent observations for stars with masses greater than 0.60 $M_\odot$. For this database no relationship is found between $Z$ and Age.

\begin{figure}[!t]
\begin{minipage}[t]{1\linewidth}
\centering
\includegraphics[width=7.8cm, trim=5.0cm 0.5cm 0cm 0cm]{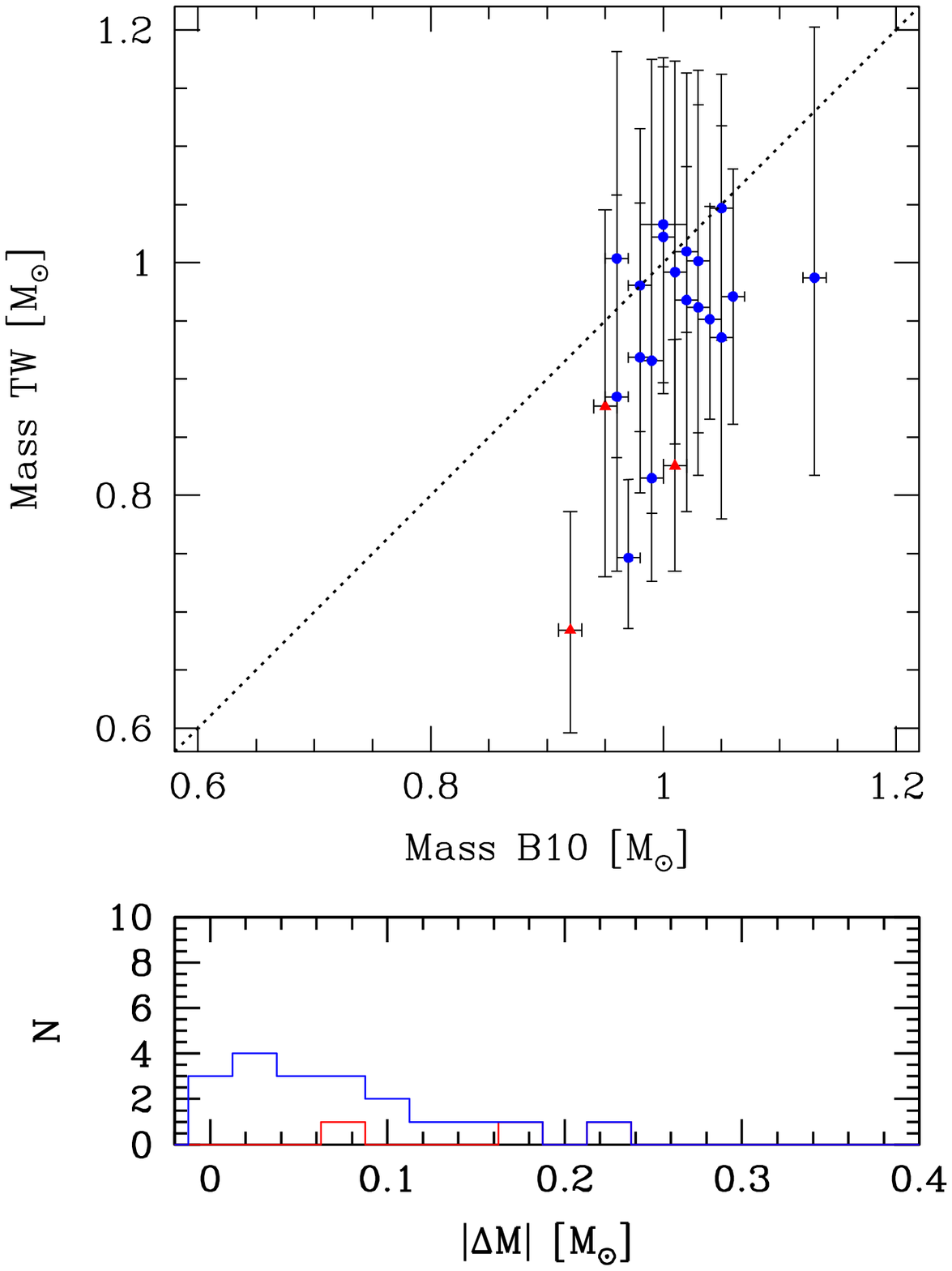}
\end{minipage}
\begin{minipage}[b]{1\linewidth}
\centering
\includegraphics[width=7.8cm, trim=5.0cm 0.5cm 0cm -1cm]{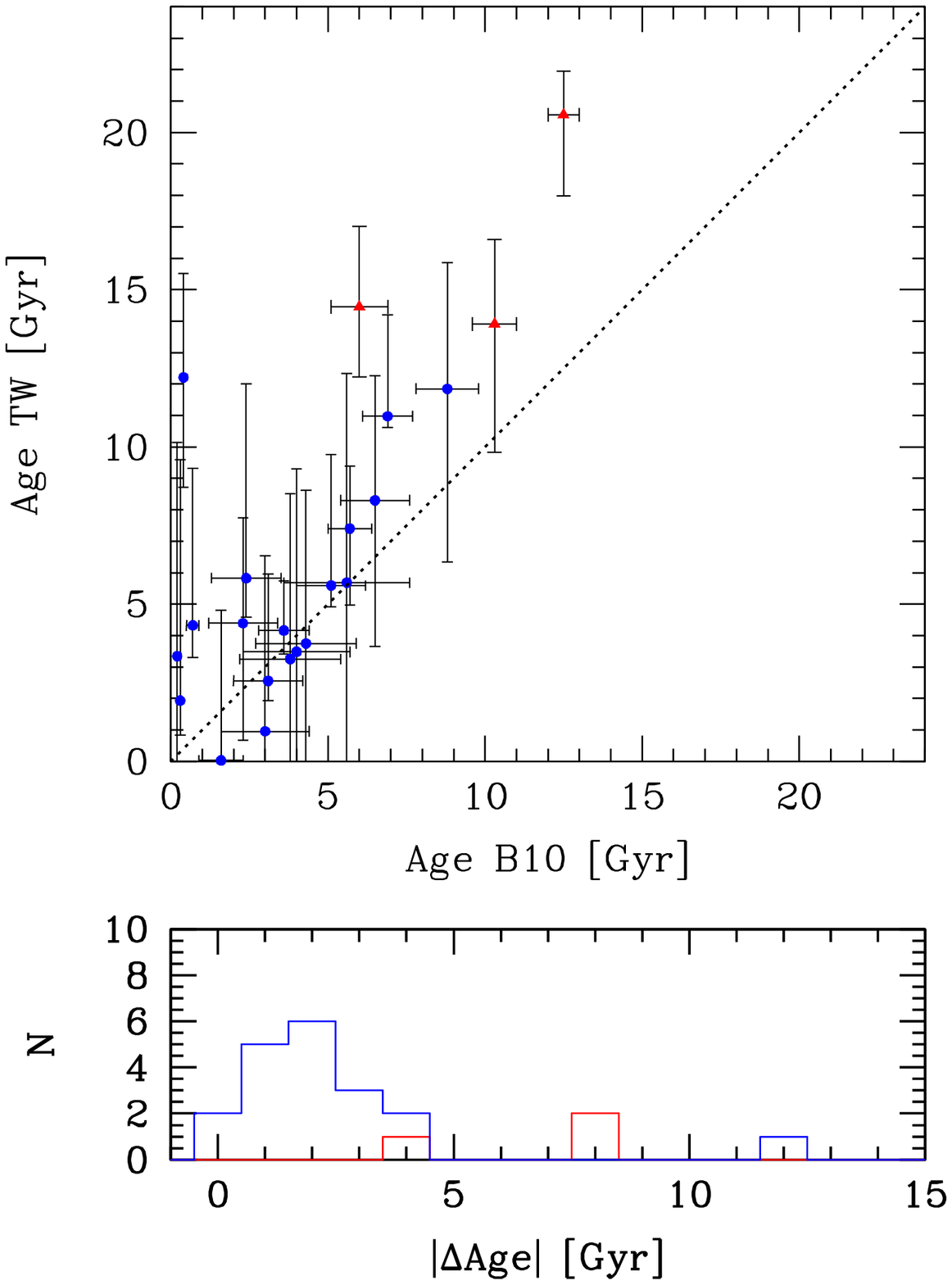}
\end{minipage}
      \caption{Comparison between masses (upper panel) and ages (lower panel), determined using theoretical models that follow a $Y$--$Z$ relation (B10) and those with non-fixed helium abundance (this work, TW). Dotted lines indicate a ratio of 1:1. Below each figure is a histogram depicting the absolute differences between both approaches.}
      \label{FigB10vsV12}
 \end{figure}

In Fig. \ref{FigB10vsV12} the masses and ages of stars determined in this study are compared to those determined by B10\footnote{Although, in principle, it is not possible to directly compare the results of this study ($M$ and Age) with the respective values determined by B10 since their masses and ages are obtained using the Y$^2$ isochrones \citep[][]{Yi_etal2001}, substantial differences are not expected, in view of the reasonable level of agreement between Y$^2$ and BaSTI found in \citet{Pietrinferni_etal2004}, and between BaSTI and PGPUC in \citet{Valcarce_etal2012}.}. For B10 a $Y$--$Z$ relation with $\Delta Y/\Delta Z=2.0$ is assumed for all stars \citep{Yi_etal2003}. This comparison demonstrates the consequences of determining masses and ages when assuming that all stars have the same $Y$ value for a given $Z$ value. Thus, we can conclude that if a universal $Y$--$Z$ relation is assumed valid for all stars, it is possible to identify differences of $|\Delta M|\approx  0.2\,M_\odot$ and $|\Delta{\rm Age}|\approx$ 2 Gyr with respect to the case when $Y$ is variable.

\subsection{Is $\Delta Y/\Delta Z$ constant?}

As pointed out in Sect. \ref{intro}, the $Y$--$Z$ relation is mostly regulated by the constant value of $\Delta Y/\Delta Z$. However, several authors have emphasized the possibility that different stellar populations might have different $Y$--$Z$ relations, implying that $\Delta Y/\Delta Z$ is not constant \citep[e.g.,][]{Catelan_deFreitasPacheco1996, Catelan2009, Nataf_Udalski2011, Nataf_etal2011,Nataf_etal2011a,Nataf_Gould2012}. To solve this conundrum, an attempt should be made to determine the $Y$ value for each star individually, as performed in this study. However, as shown in Fig. \ref{FigErrors}, this method requires highly accurate measurements of surface gravity, with $\sigma_{\log g}\lesssim0.05$ dex, otherwise large error bars may make it impossible to obtain a reliable conclusion. In the near future, high-precision observations and more accurate theoretical models could establish whether $\Delta Y/\Delta Z$ is not constant only in some GCs, or if this is also valid among the field populations in galaxies.

\subsection{Consequences of non-canonical effects}

Given that the surface gravity is used here as one of the main parameters to determine the FSPs, it is important to know which stellar properties may potentially affect the measurement of $g$. Recently \citet[][hereafter B12]{Basu_etal2012} have tested how properties of single stars ($M$, $R$, and $\log g$) are affected by the uncertainties in stellar models. They studied the effects induced by different metallicity scales, atmospheric models ($T-\tau$ relation), mixing length parameters $\alpha_l$, and shifts in the temperature scale for observed stars with and without seismic data. We focus on the last, since our stars also lack seismic data. In this case, their results show that the errors in the determination of $M$ and $R$ are around 8\% and 14\%, respectively, for errors in the observational data of $\sigma_{T_{\rm eff}}=50$ K, $\sigma_{\rm [Fe/H]}=0.1$ dex, and $\sigma_{\log g}=0.1$ dex. 

While adopting variable $\alpha_l$ values, as suggested by B12, represents an interesting means of improving the estimations of the stellar parameters \citep[although in this case, care should be taken to properly calibrate the temperature scale implied by the models, so that it retains some predictive power for stars other than the Sun, which at present remains the sole calibrator that is used to obtain $\alpha_l$; see, e.g.,][and references therein]{Catelan2012}, a variable initial helium abundance represents another potentially feasible alternative, especially at the main sequence and sub giant branch phases. Regarding this topic and because here and in B12 non-rotating stellar models are used, one can speculate that some calibration parameters of the model \citep[such as $\alpha_l$,][]{Kapyla_etal2005} may indeed not be valid for all stars, since stellar rotation may differ from the rotation pattern which is observed in the Sun \citep[e.g.,][]{deMedeiros_etal2006,Cortes_etal2009}~-- although in practice the deviations from the canonical case may be small.

The interested reader is also referred to Sect. 5 in \citet{Gennaro_etal2010} for a detailed discussion of possible sources of uncertainties that may be present in comparisons of this kind.

Since one of the aims of this paper is to determine whether the $Y$--$Z$ relation can be used with stellar models with the same characteristics as those created by the PGPUC SEC, an analysis of these possible differences is beyond the scope of this paper. Nevertheless, we emphasize the importance of taking $Y$ into account as an unknown parameter, when attempting to derive the FSPs of stars.

\section{Conclusions}
\label{Conclusions}

In this paper a new method for determing the FSPs ($Y$, $Z$, $M$, and Age) of nearby stars is presented, where the required input parameters include the chemical composition ([Fe/H], as well as $[\alpha /{\rm Fe}]$), $M_{\rm bol}$ or the absolute magnitude in a given filter (e.g., $M_V$), $T_{\rm eff}$, and $\varg$. 

This method uses almost 600 new evolutionary tracks for 7 masses ($0.5\le M/M_\odot\le1.1$), 7 helium abundances ($0.230\le Y \le0.370$), and 12 metallicities ($1.6\times 10^{-4}\le Z \le 6.0\times 10^{-2}$) for $[\alpha/{\rm Fe}]=0.0$. These new evolutionary tracks are now included in the PGPUC online database, making it possible to interpolate evolutionary tracks and isochrones with a variable $\alpha$--element distribution ($0.0 \le [\alpha/{\rm Fe}] \le +0.3$).

This method is tested using the C06 database, concluding that an age of 5 Gyr for all nearby stars is not a good approximation, since this assumption leads to an increasing underestimation of $Y$ as metallicity decreases. Moreover, the method is only reliable for masses higher than about 0.60 $M_\odot$, in view of problems in determining the radius of stars below that limit.

The FSPs are determined for 20\% of the stars of the B10 spectroscopic database. When comparing the masses and ages obtained in this study (where $Y$ is not assumed a priori to depend on $Z$) with those obtained by B10 (where $Y$ follows a $Y$--$Z$ relation with $\Delta Y/\Delta Z=2.0$), average differences in the masses and ages of $|\Delta M|\approx0.02\,M_\odot$ and $|\Delta{\rm Age}|\approx$ 2 Gyr are recorded, respectively, for the range of masses used here ($0.5\le M/M_\odot\le1.1$).

\subsection*{Acknowledgments}
We thank the anonymous referee for her/his comments, which have helped improve this paper. Support for A.A.R.V. and M.C. is provided by the Ministry for the Economy, Development, and Tourism's Programa Iniciativa Cient\'{i}fica Milenio through grant P07-021-F, awarded to The Milky Way Millennium Nucleus; by Proyecto Basal PFB-06/2007; by Proyecto FONDECYT Regular \#1110326; and by Proyecto Anillo de Investigaci\'{o}n en Ciencia y Tecnolog\'{i}a PIA CONICYT-ACT 86. A.A.R.V. and J.R.M. acknowledge additional support from CNPq, CAPES, and INEspa\c{c}o agencies.

\bibliographystyle{aa}
\bibliography{avalcarce}

\end{document}